\documentclass[showpacs,amsmath,amssymb,prb,preprint]{revtex4}
\usepackage{graphicx}% Include figure files
\usepackage{dcolumn}% Align table columns on decimal point
\usepackage{bm}% bold math

\begin{document}

%\preprint{APS/123-QED}

\title{Spin-signal propagation in time-dependent noncollinear spin transport}

\author{Yao-Hui Zhu}\email{yaohuizhu@gmail.com}
\author{Burkard Hillebrands}
\author{Hans Christian Schneider}
\homepage{http://www.physik.uni-kl.de/schneider}
\affiliation{Physics Department and Research Center OPTIMAS,
University of Kaiserslautern, 67653 Kaiserslautern,
Germany}


\begin{abstract}
Using a macroscopic analysis, we demonstrate that time-dependent noncollinear
spin transport may show a wavelike character. This leads to
modifications of pure spin-diffusion dynamics and allows one to
extract a finite spin-signal propagation velocity. We numerically
study the dynamics of a pure spin current pumped into a nonmagnetic
layer for precession frequencies ranging from GHz to THz.
\end{abstract}

\maketitle

\section{Introduction\label{sec1}}

Transporting information encoded in electronic spins through layers of
ferromagnetic and normal metals is a central theme of
magnetoelectronics.~\cite{Bra06} Structures, in which all spins are
are essentially \emph{collinear}, i.e., parallel or antiparallel, have
been thoroughly investigated in experimental and theoretical
studies. The quasi-static properties for the special case of
structures with collinear spin and magnetization directions where the
spin-polarized current flows perpendicularly to the plane of the
layers,~\cite{Pra91} can be analyzed in terms of a scalar
space-dependent spin accumulation for up and down
spins.~\cite{vanSon87,vf} The functionality of collinear
magnetoresistive structures can be enhanced by including tunneling
elements.~\cite{Sch00,Rash00,fj01} Although collinear spin transport
is of importance for certain variants of giant and tunneling
magnetoresistance effects, a \emph{non-collinear} alignment of spin
and magnetization orientations leads to additional degrees of freedom
for the manipulation of spin angular momentum and has attracted much
attention in recent years.~\cite{Tser05,Bra06} For instance, one can
exploit the angular dependence of the giant magnetoresistance
effect~\cite{barnas:prb97}, or can change the the alignment of spins
by spin currents, leading to the phenomenon of spin transfer
torque~\cite{Slon96,Ber96,Tsoi98,Myers99} and potential novel
applications.~\cite{Katine:jmmm07}. A different method to exploit the
freedom of noncollinear spin orientations in magnetic multilayers is
the use of magnetization precession in a ferromagnetic layer, which
``pumps'' a spin currents into an adjacent nonmagnetic
metal.~\cite{Tser02} A precessing magnetization, which is necessary
for spin pumping, creates the need to deal with a time-dependent
orientation of the spins in the whole multilayer, so that it becomes
essential to study dynamical noncollinear spin transport problems.

We are concerned with a theoretical analysis of the propagation of
signals encoded in a spin current, which flows through a multilayer
structure with noncollinear magnetization and spin directions. Most
investigations of time-dependent noncollinear spin transport are based
on the Bloch-Torrey diffusion equations for the nonequilibrium
magnetization or spin accumulation.~\cite{Zut04} These equations
essentially describe spin transport as a diffusion process and
therefore show the same problem as the spin diffusion equation for
collinear spins~\cite{zhang02,Rash02,zhang05,cy06}: no finite
propagation velocity for a spin signal can be defined because the
diffusion equation leads to a finite spin current density everywhere
as soon as there is a source. Recently, we showed that this difficulty
can be resolved for collinear spin transport by using a ``telegraph''
equation, which generalizes the diffusion equation, and leads to
noticeable differences from the diffusion equation results for
frequencies exceeding several 100 GHz for metals such as
copper.~\cite{Zhu08} Importantly, the telegraph equation shows a
wave-diffusion duality, which enables one to define a finite
propagation velocity for the spin signal. In this paper, we use a
similar treatment for noncollinear spin transport to show how a finite
signal propagation velocity arises in this case.  We predict that
noncollinear spin transport at high frequencies shows a dynamics that
is more complicated than what is expected from an ana\-lysis using the
spin-diffusion equation. We numerically analyze the propagation of a
spin current pumped into a nonmagnetic metal by a precessing
magnetization in an adjacent ferromagnetic layer.

This paper is organized as follows. In Sec.~\ref{sec2}, we present the
macroscopic dynamical equations governing noncollinear spin
transport. In Sec.~\ref{sec3}, the dynamical equations are combined
into a telegraph equation, which is studied analytically to discuss
qualitative aspects of dynamical noncollinear spin-transport. In
Sec.~\ref{sec4}, we solve numerically the dynamical equations for the
spin transport, and the main conclusions are summarized in
Sec.~\ref{sec5}.

\section{Dynamical equations\label{sec2}}

In nonmagnetic conductors and some ferromagnetic metals,~\cite{Zha04}
the dynamics of conduction electrons under the influence of external
fields can be described by a generalized semiclassical Boltzmann
equation~\cite{Smi89,Ram86}
\begin{equation}\label{bte1}
i\hbar \frac{\partial\hat\rho}{\partial{t}}
+\frac{i}{2}\left\{\frac{\partial\hat\varepsilon}{\partial\vec{k}},
\frac{\partial\hat\rho}{\partial\vec{r}}\right\}
-\frac{i}{2}\left\{\frac{\partial\hat\varepsilon}{\partial\vec{r}},
\frac{\partial\hat\rho}{\partial\vec{k}}\right\}
=[\hat\varepsilon,\hat\rho]
+i\hbar\frac{\partial\hat\rho}{\partial{t}}\bigg|_{\mathrm{col}},
\end{equation}
which we take as the starting point for our analysis of time-dependent
noncollinear electron-spin transport in these systems.  In
Eq.~\eqref{bte1}, $\hat\rho(\vec{r},\vec{k},t)$ is the single particle
density matrix in spin space,
\begin{equation}\label{matrix}
\hat\rho= \left(
  \begin{array}{cc}
    \rho_{\uparrow\uparrow} & \rho_{\uparrow\downarrow} \\
    \rho_{\downarrow\uparrow} & \rho_{\downarrow\downarrow} \\
  \end{array}
\right),
\end{equation}
$\hat\varepsilon(\vec{r},\vec{k},t)$ is the effective single-particle
energy matrix, and $\{\cdot,\cdot\}$ and $[\cdot,\cdot]$ denote
respectively the anticommutator and commutator for matrices in spin
space. For completeness, we remark that in Eq.~(\ref{matrix}), the
single-particle density matrix
\begin{equation}
\rho_{ss'}(\vec{r},\vec{k},t)=\frac{V}{(2\pi)^3}\int d^3q
e^{i\vec{q}\cdot\vec{r}} \langle {c}_{\vec{k}-\vec{q}/2,s'}^{\dag}
c_{\vec{k}+\vec{q}/2,s}\rangle,
\end{equation}
is defined by a statistical average over creation and annihilation
operators $c^{\dag}$ and $c$, with normalization volume $V$. The
diagonal matrix elements $\rho_{\uparrow\uparrow}$ and
$\rho_{\downarrow\downarrow}$ are the electron distribution functions
of the spin-up and spin-down, respectively, whereas the off-diagonal
elements $\rho_{\uparrow\downarrow}=\rho_{\downarrow\uparrow}^\ast$
represent the spin coherence.~\cite{sham:mmm99}
%
%\cite{sham:mmm99,Che07,Krau08}
%
Because the unit matrix $\hat{I}$ and the Pauli matrices
$\hat\sigma_x$, $\hat\sigma_y$, $\hat\sigma_z$ form a basis for
$2\times2$ matrices, the spin-density matrix $\hat\rho$ can be
represented by
$\hat\rho=(1/2)[(\rho_{\uparrow\uparrow}+\rho_{\downarrow\downarrow})
  \hat{I}+\vec{u}\cdot\hat{\bm{\sigma}}]$, where
$\vec{u}=\mathrm{Tr}(\hat{\bm{\sigma}}\hat\rho)
=(2\mathrm{Re}\rho_{\uparrow\downarrow},-2\mathrm{Im}\rho_{\uparrow\downarrow},
\rho_{\uparrow\uparrow}-\rho_{\downarrow\downarrow})$ is the Bloch
vector and $\hat{\bm{\sigma}}$ the vector of Pauli matrices.

Before proceeding from Eq.~\eqref{bte1} for the spin-density matrix to
equations for macroscopic quantities, such as spin currents and spin
accumulation, we list a few assumptions made about quantities
occurring in Eq.~\eqref{bte1}. First, we consider only layered
structures whose extensions perpendicular to the growth direction ($x$
axis) are infinite, and we also assume that the electric fields
$\vec{E}=E\vec{x}/|\vec{x}|$ is oriented along the growth
direction~$x$. Second, the effect of magnetic fields on the orbital
motion of electrons is neglected. These magnetic fields include the
static external magnetic field $\vec{B}_s$ and the magnetic field
generated by induction due to the time-dependent electric field
$\vec{E}(x,t)$.~\cite{Ash76} We therefore assume that the electric
field $E(x,t)=-\partial\phi(x,t)/\partial x$ can be derived from a
time-dependent electric potential $\phi(x,t)$.  Third, an isotropic
effective mass model for the spin-degenerate condutcion electrons is
used, i.e., $\varepsilon_k=\hbar^2k^2/(2m^\ast)=m^{\ast}v^2/2$, where
$\vec{k}$ and $\vec{v}$ denote the the electron wave vector and
velocity, respectively. Thus we have to deal with a spin density
matrix $\hat\rho$ that depends only on $x$ and has cylindrical
symmetry around the $x$ axis in $k$ space.

Finally, we make a relaxation-time approximation for the collision
term~\cite{Qi03}
\begin{equation}\label{rta}
\frac{\partial\hat\rho}{\partial{t}}\bigg|_{\mathrm{col}}=
-\frac{\hat\rho-\langle\hat\rho\rangle_a}{\tau}
-\frac{\langle\hat\rho\rangle_a
-(\hat{I}/2)\mathrm{Tr}\langle\hat\rho\rangle_a}{T_1},
\end{equation}
where $\tau$ and $T_1$ are the momentum and spin relaxation times,
respectively. Moreover, $\langle\hat\rho\rangle_a\equiv(4\pi)^{-1}
\int{d}\Omega_{\vec{k}}\hat\rho$ is the angular average in the
momentum space. By using Eq.~(\ref{rta}) for the collision term, we
have assumed that the longitudinal spin relaxation time $T_1$ is
equal to the transverse one $T_2$. The validity of this
approximation is discussed in detail by Ref.~\onlinecite{Zut04}.
Note that $T_1$ in Eq.~(\ref{rta}) is one half of
$\tau_{\mathrm{sf}}$ used in Eq.~(2) of Ref.~\onlinecite{Qi03}.

With above simplifications, the effective single-particle energy
$\hat\varepsilon(\vec{r},\vec{k},t)$ is simplified to
$\hat\varepsilon(x,|\vec{v}|,t)=\varepsilon_0\hat{I}+\hat\varepsilon_s$,
where $\varepsilon_0=\hbar^2 k^2/(2m^{\ast})-e\phi(x,t)$ and
$\hat\varepsilon_s=-\bm{\mu}\cdot\vec{B}_s
=\mu_B\bm{\sigma}\cdot\vec{B}_s$. Therefore, Eq.~(\ref{bte1})
simplifies to
\begin{equation}
\label{bte2}
\frac{\partial\hat\rho}{\partial{t}}
+v_x\frac{\partial\hat\rho}{\partial{x}}
-\frac{eE}{m^\ast}\frac{\partial\hat\rho}{\partial{v}_x}
+\frac{1}{2}\gamma(\vec{u}\times\vec{B}_s)\cdot\bm{\sigma}
=-\frac{\hat\rho-\langle\hat\rho\rangle_a}{\tau}
-\frac{\langle\hat\rho\rangle_a
-(\hat{I}/2)\mathrm{Tr}\langle\hat\rho\rangle_a}{T_1},
\end{equation}
where $\gamma=g\mu_B/\hbar$ is the absolute value of the electron
($g\approx2$) gyromagnetic ratio.

To derive macroscopic spin transport equations comparable with the
Bloch-Torrey diffusion equation, we need to sum over the electron wave
vector $\vec{k}$ or, equivalently, the velocity $\vec{v}$ in
Eq.~(\ref{bte2}). We first derive an equation for the spin
density~\cite{Qi03,Mar94} by multiplying both sides of
Eq.~(\ref{bte2}) by $\hat{\bm{\sigma}}/V$, taking the trace, and
summing over $\vec{v}$
\begin{equation}
\frac{\partial\vec{n}_s(x,t)}{\partial{t}}
=-\gamma\vec{n}_s(x,t)\times\vec{B}_s
-\frac{\vec{n}_s(x,t)}{T_{1}}
-\frac{\partial\vec{\jmath}_{s}(x,t)}{\partial{x}},\label{spinsum8}
\end{equation}
where $\vec{n}_s(x,t)=V^{-1}\sum_{\vec{v}}
\mathrm{Tr}(\bm{\sigma}\hat\rho)=V^{-1}\sum_{\vec{v}}\vec{u}$ and
$\vec{\jmath}_{s}(x,t)=V^{-1}\sum_{\vec{v}}v_{x}
\mathrm{Tr}(\bm{\sigma}\hat\rho)=V^{-1}\sum_{\vec{v}}v_{x}\vec{u}$
are the spin density and spin current density, respectively.  For the
spin current density, we multiply both sides of Eq.~(\ref{bte2}) by
$v_x\hat{\bm{\sigma}}/V$, take the trace, and sum over
$\vec{v}$. Using the expansion~\eqref{Legendre} for the velocity
dependence of the spin density matrix and the procedure in Appendix~A,
we obtain
\begin{equation}
\vec{\jmath}_{s}(x,t)=-D\frac{\partial\vec{n}_s(x,t)}{\partial{x}}
-\mu{E}(x,t)\vec{n}_s(x,t)
-\tau\gamma\vec{\jmath}_{s}(x,t)\times\vec{B}_s
-\tau\frac{\partial\vec{\jmath}_{s}(x,t)}{\partial{t}},\label{current9}
\end{equation}
where 
\begin{equation}
D = \frac{v^2_{\mathrm{F}}}{3}\tau
\label{diff-const}
\end{equation}
is the diffusion constant and $\mu=e\tau/m^{\ast}$ the electron
mobility. Note that $\vec{n}_s(x,t)$ and $\vec{\jmath}_{s}(x,t)$ defined
above are the particle (electron) number densities, which can be
converted to the charge, spin, and magnetic moment densities by
multiplication with $-e$, $\hbar/2$, and $-\mu_{B}$, respectively. The
spin density $\vec{n}_s(x,t)$ can also be converted to the chemical
potential difference $\bm{\mu}_s(x,t)$, i.e., the spin accumulation,
by the relation $\vec{n}_s(x,t)=\mathcal{N}\bm{\mu}_s(x,t)$, where
$\mathcal{N}=4\pi{m}^{\ast2}v_F/h^3$ is the density of states at the
Fermi level of the electron gas for one spin
orientation.~\cite{Tser02b}

Equation~\eqref{current9} resembles the dynamical equation for the
spin current derived by Qi and Zhang~\cite{Qi03} using a ``mean
field'' approximation. Our derivation shows that their quantity
$\overline{v_{x}^{2}}$ is equal to $v^2_{F}/3$. As will be discussed
in the next section, this is the wavefront velocity for a spin
disturbance, which plays an important role in spin-signal propagation
dynamics~\cite{Zhu08}.

\section{Telegraph equation\label{sec3}}

To see the physical significance of Eqs.~(\ref{spinsum8}) and
(\ref{current9}) for the time-dependent noncollinear spin transport
and compare them with the Bloch-Torrey equation, we combine them by
eliminating $\vec{\jmath}_s(x,t)$ into a form reminiscent of a
telegraph equation\cite{Zhu08}
\begin{equation}\label{tele}
\begin{split}
&\frac{\partial^{2}\vec{n}_s(x,t)}{\partial{t}^{2}}+(\frac{1}{\tau}
+\frac{1}{T_{1}})\frac{\partial\vec{n}_s(x,t)}{\partial{t}}
+\frac{\vec{n}_s(x,t)}{\tau{T}_{1}}
+\gamma\left[2\frac{\partial}{\partial{t}}
+(\frac{1}{\tau}+\frac{1}{T_{1}})\right]\vec{n}_s(x,t)\times\vec{B}_s\\
&\qquad+\gamma^{2}[\vec{n}_s(x,t)\times\vec{B}_s]\times\vec{B}_s\\
&=c_s^{2}\frac{\partial^{2}\vec{n}_s(x,t)}{\partial{x}^{2}}
+\frac{\mu{E}(x,t)}{\tau}\frac{\partial\vec{n}_s(x,t)}{\partial{x}}
+\frac{\mu}{\tau}\frac{\partial{E}(x,t)}{\partial{x}}\vec{n}_s(x,t).
\end{split}
\end{equation}
Similarly, one can also derive a telegraph equation for
$\vec{\jmath}_s(x,t)$ by eliminating $\vec{n}_s(x,t)$ from
Eqs.~(\ref{spinsum8}) and (\ref{current9}). Equation~(\ref{tele})
contains a second-order time derivative, which is absent in the spin
diffusion equation. The second-order time and space derivatives lead
to a wave character in addition to its diffusion character, and thus
yield a well-defined propagation velocity $c_s$ for the signal in
time-dependent noncollinear spin transport in a similar way to the
collinear case.~\cite{Zhu08}

Assuming the static magnetic field $\vec{B}_s$ to be oriented along
the $z$ axis and separating the components perpendicular (transverse)
and parallel (longitudinal) to $\vec{B}_s$ in Eq.~(\ref{tele}), we
have
\begin{align}\label{telex}
\frac{\partial^{2}n_s^{x(y)}}{\partial{t}^{2}}
+\left(\frac{1}{\tau}+\frac{1}{T_{1}}\right)
\frac{\partial{n}_s^{x(y)}}{\partial{t}}
+\frac{n_s^{x(y)}}{\tau{T}_{1}}
& +(-)\gamma{B}_s\left[2\frac{\partial}{\partial{t}}
+\left(\frac{1}{\tau}+\frac{1}{T_{1}}\right)\right]n_s^{y(x)}
-\gamma^{2}B_s^{2}n_s^{x(y)} \nonumber \\
&\mbox{}=c_s^{2}\frac{\partial^{2}n_s^{x(y)}}{\partial{x}^{2}}
+\frac{\mu E}{\tau}\frac{\partial{n}_s^{x(y)}}{\partial{x}}
+\frac{\mu}{\tau}\frac{\partial E}{\partial{x}}n_s^{x(y)}
\end{align}
and
\begin{equation}
\frac{\partial^{2}n_s^{z}}{\partial{t}^{2}}
+\left(\frac{1}{\tau}+\frac{1}{T_{1}}\right)
\frac{\partial{n}_s^{z}}{\partial{t}}+\frac{n_s^{z}}{\tau{T}_{1}}
=c_s^{2}\frac{\partial^{2}n_s^{z}}{\partial{x}^{2}}
+\frac{\mu E}{\tau}\frac{\partial{n}_s^{z}}{\partial{x}}
+\frac{\mu}{\tau}\frac{\partial E}{\partial{x}}n_s^{z}.
\end{equation}
In the following, only the equation for the transverse component
[Eq.~(\ref{telex})] will be discussed, since the equation for the
longitudinal component is similar to that of the collinear
case.~\cite{Zhu08} For vanishing electric field, i.e., $E=0$, we seek
damped and dispersive wave solutions to Eq.~(\ref{telex}) of the form
\begin{eqnarray}
n_s^{x}(x,t)&=&n_{0}\exp[i(kx-\omega{t})],\label{solutionx}\\
n_s^{y}(x,t)&=&n_{0}\exp[i(kx-\omega{t}+\phi)],\label{solutiony}
\end{eqnarray}
where $\omega$ is the angular frequency and $k=k_{r}+ik_{i}$ the
complex wave vector. Substituting Eqs.~(\ref{solutionx}) and
(\ref{solutiony}) into Eqs.~(\ref{telex}), we obtain the dispersion
relation
\begin{equation}\label{dis1}
\omega^{2}+i\omega(1/\tau+1/T_{1})-1/(\tau{T}_{1})
-c_s^{2}k^{2}+\gamma^2B_s^2
-\gamma{B}_s\left[2\omega+i\left(1/\tau+1/T_{1}\right)\right]\sin\phi=0,
\end{equation}
where $\phi$ is restricted to $\phi=\pm(\pi/2)+2n\pi$ and $n$ is an
integer, because $n^x_s$ and $n^y_s$ must satisfy the system of
equations~(\ref{telex}) at the same time. According to
Eqs.~(\ref{solutionx}) and (\ref{solutiony}), $\phi=+(-)\pi/2$
corresponds to the rotation direction of the transverse component of
$\vec{n}_{s}(x,t)$ with $x$ at time $t$. For definiteness, we study the
case with $\phi=\pi/2$ in the following. Substituting $k=k_{r}+ik_{i}$
into Eq.~(\ref{dis1}) and separating the real and imaginary parts, we
have
\begin{eqnarray}
k_{r(i)}^2=\frac{1}{2c_s^2}\left[\sqrt{b^{2}+
\omega_{\mathrm{eff}}^{2}\alpha^{2}}+(-)b\right],
\end{eqnarray}
where $\omega_{\mathrm{eff}}=\omega-\gamma{B}_s$ and
$b=\omega_{\mathrm{eff}}^{2}-\xi$. Here, the constants
$\alpha=1/\tau+1/T_{1}$ and $\xi=1/(\tau{T}_{1})$ have been
introduced. The wavelength and damping length can be defined as
$\lambda=2\pi/k_{r}$ and $l_d=1/k_{i}$, respectively. The equation
of the critical angular frequency $\omega_{\mathrm{crit}}$, above
which the wave character is significant, can be derived by setting
$\lambda=l_{d}$,
\begin{equation}\label{critical}
\omega_{\mathrm{eff}}^{\mathrm{crit}}\tau =\frac{1}{2}\left[
\delta(1+\eta)+\sqrt{\delta^{2}(1+\eta)^{2}+4\eta}\right]
\approx\delta+(\delta+\frac{1}{\delta})\eta,
\end{equation}
where $\delta=\pi-1/(4\pi)\approx3.06$ and $\eta=\tau/T_1$. Then, we
have $\omega_{\mathrm{crit}}\tau=3.06+3.4\eta+\tau\gamma{B}_s$
approximately.

\section{Dynamics of pumped spin current\label{sec4}}

In this section, we study the evolution of the spin current injected
into a nonmagnetic layer by the spin-pumping mechanism.~\cite{Tser02}
In a junction composed of a ferromagnetic ($x<0$) and a nonmagnetic
($x>0$) layer, the magnetization precession of the ferromagnet
around an external magnetic field $\vec{B}_{\text{pump}}$ acts as a ``spin
pump'' which transfers spin angular momentum from the ferromagnet to
the adjacent nonmagnetic layer. The spin current density pumped into
the nonmagnetic layer is\cite{Tser02,Bra02,Tser02b}
\begin{equation}\label{pump}
\vec{\jmath}_{s}^{\mathrm{pump}}=\frac{1}{2\pi}
\frac{g^{\uparrow\downarrow}}{S}\vec{m}\times\frac{d\vec{m}}{dt},
\end{equation}
where $g^{\uparrow\downarrow}$ is the spin-mixing conductance and
$S$ the area of the interface. Here, $\vec{m}$ is the unit vector
for the magnetization of the ferromagnet. Note that the pumped spin
current has been converted to a particle number current density
$\vec{\jmath}_{s}^{\mathrm{pump}}$. Since we are interested in the
spin current pumped into a nonmagnetic layer and not in the dynamics
of the ferromagnet, we neglect the back-flow spin current
$\vec{I}_{s}^{\mathrm{back}}$, which flows from the nonmagnetic
layer to the ferromagnet due to the spin accumulation in the
nonmagnetic layer.~\cite{Tser02b} Although the back-flow spin
current can limit the achievable spin current into the nonmagnetic
conductor, we do not approach this limit here. With this
simplification, we have
$\vec{\jmath}_{s}^{\mathrm{pump}}=\vec{\jmath}_s(x=0,t)$, where
$\vec{\jmath}_s(x=0,t)$ is the spin current density at the left
boundary of the nonmagnetic layer. Separating the components
perpendicular and parallel to the magnetic field $\vec{B}_{\text{pump}}$, we
can write $\vec{\jmath}_{s}(x=0,t)$ as
\begin{eqnarray}\label{boundary}
j_s^x(x=0,t)&=&g^{\uparrow\downarrow}(4\pi{S})^{-1}\omega\sin(2\theta)\cos(\omega{t})\\
j_s^y(x=0,t)&=&g^{\uparrow\downarrow}(4\pi{S})^{-1}\omega\sin(2\theta)\sin(\omega{t})\\
j_s^z(x=0,t)&=&g^{\uparrow\downarrow}(2\pi{S})^{-1}\omega\sin^2\theta,
\end{eqnarray}
where $\omega$ is the angular frequency of both the magnetization
precession and the spin current density $\vec{\jmath}_{s}(x=0,t)$.  Here,
$\omega{t}$ is the angle between $\vec{\jmath}_s^{\bot}$ ($j_s^x$ and
$j_s^y$) and the $x$-axis. $\theta$ is the angle between $\vec{m}$ and
$\vec{B}_{\text{pump}}$, and meanwhile $\theta$ is also the angle
between $\vec{\jmath}_{s}(x=0,t)$ and $xy$-plane. The amplitude of
$\vec{\jmath}_s^{\bot}$ is much larger than $j_s^z$, since $\theta$ is very
small under the usual radio-frequency excitation
conditions.~\cite{Bra02} Therefore, we will focus on
$\vec{\jmath}_s^{\bot}$ in the following.

The propagation of $\vec{\jmath}_{s}^{\bot}(x=0,t)$ into the
nonmagnetic layer is described by Eqs.~(\ref{spinsum8}) and
(\ref{current9}). In a typical setup for spin pumping, there is no
electric or magnetic field in the nonmagnetic layer, i.e., $E=0$
and $\vec{B}_s=0$. Now, separating the components perpendicular
and parallel to the magnetic field $\vec{B}_{\text{pump}}$, we can rewrite
Eqs.~(\ref{spinsum8}) and (\ref{current9}) as
\begin{eqnarray}
\frac{\partial{n}_s^{+}}{\partial{t}}+\frac{\partial{j}_{s}^{+}}{\partial{x}}=
-\frac{n_s^{+}}{T_{1}},\label{simxy1}\\
j_{s}^{+}=-D\frac{\partial{n}_s^{+}}{\partial{x}}
-\tau\frac{\partial{j}_{s}^{+}}{\partial{t}},\label{simxy2}
\end{eqnarray}
where $n_s^{+}=n_s^{x}+in_s^{y}$ and
$j_{s}^{+}=j_{s}^{x}+ij_{s}^{y}$ are introduced to simplify the
notations. The equations for the parallel component can be obtained
after replacing $n_s^+$ and $j_s^+$ by $n_s^z$ and $j_s^z$ in
Eqs.~(\ref{simxy1}) and (\ref{simxy2}), respectively. The method of
characteristics used for the numerical solution to
Eqs.~(\ref{simxy1}) and (\ref{simxy2}) is outlined in
Appendix~\ref{numerical}.

\begin{figure}
\includegraphics[width=0.4\textwidth]{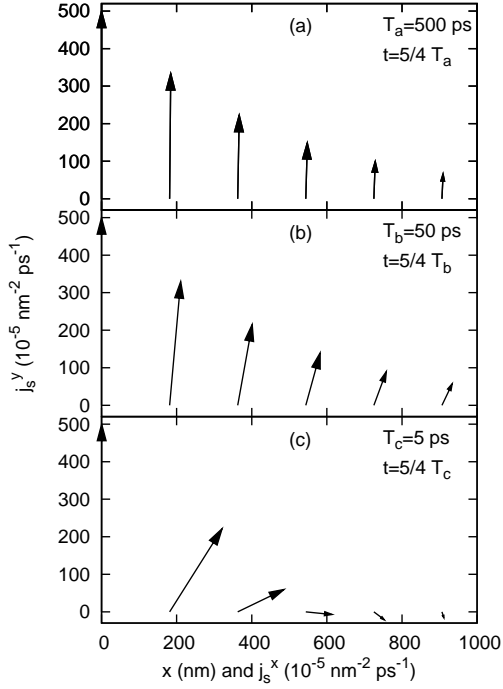}
\caption{Snapshots of the spin current density $\vec{\jmath}_s^{\bot}$
at $t=5/4$~$T_a$, $5/4\ T_b$, and $5/4\ T_c$, for the frequencies,
$\nu_a$, $\nu_b$, and $\nu_c$, respectively (see text).
$\vec{\jmath}_s^{\bot}$ is plotted as vector starting from its
$x$-coordinate. \label{fig1}}
\end{figure}

In our numerical calculation, Cu and permalloy (Py) are chosen as the
materials for the nonmagnetic and ferromagnetic layers,
respectively. The Fermi velocity of Cu is $v_F=1570$\,nm/ps and thus
the wave-front velocity is $c_s=v_F/\sqrt{3}=906$\,nm/ps. The momentum
and spin relaxation times are $\tau=0.07$\,ps and $T_1=3.5$\,ps,
respectively.  The critical frequency can be estimated to be
$\nu_{\mathrm{crit}}=\omega_{\mathrm{crit}}/(2\pi)=7.11$\,THz from
Eq.~(\ref{critical}). We study several pumping frequencies:
$\nu_a=1/T_a=2$\,GHz, $\nu_b=1/T_b=20$\,GHz, $\nu_c=1/T_c=200$\,GHz, and
$\nu_d=1/T_d= 8.33$\,THz. For a Py/Cu junction,~\cite{Tser02b}
$g^{\uparrow\downarrow}S^{-1}$ is on the order of $10^{15}$
cm$^{-2}$. The precession cone angle $\theta$ can reach
$15^\circ$ for a sufficiently intense radio-frequency
field.~\cite{Bra02} Therefore, we choose the amplitude of
$\vec{\jmath}_s^{\bot}$, i.e.,
$g^{\uparrow\downarrow}(4\pi{S})^{-1}\omega\sin(2\theta)$, to be
$5\times10^{-3}$\,nm$^{-2}$\,ps$^{-1}$ for the frequencies mentioned
above.

Figure \ref{fig1} shows snapshots of the spin current density
$\vec{\jmath}_s^{\bot}$ at $t=(5/4)\,T_a$, $(5/4)\,T_b$, and $(5/4)\,T_c$, for
the frequencies, $\nu_a$, $\nu_b$, and $\nu_c$, respectively.
According to Eqs.~(\ref{boundary}), $\vec{\jmath}_s^{\bot}(x=0,t)$ points
in the direction of the $y$-axis at $t=5/4\,T_{a(b,c)}$, which can
also be seen in Fig.~(\ref{fig1}). Figure~\ref{fig1} (a) shows that
$\vec{\jmath}_s^{\bot}$ points along $y$-axis nearly at all $x$ points
except that it deviates from the $y$-axis slightly at positions far
away from $x=0$. The results in Fig.~\ref{fig1} (a) are approximately
consistent with those obtained from the diffusion equation in
Refs.~\onlinecite{Tser02b} and \onlinecite{Bra02}, where it is shown
that both the spin current and spin accumulation point along the same
direction at all positions for all frequencies at certain time point
$t$. This agreement means that the diffusion equation provides a good
description in the low frequency range.~\cite{Ger06} The deviation of
$\vec{\jmath}_s^{\bot}$ from the $y$-axis at $x>0$ increases with frequency
and becomes noticeable at $\nu_b=1/T_b=20$\,GHz as shown in
Fig.~\ref{fig1}~(b). Therefore, the applicability of the diffusion
equation is questionable in this frequency region. At even higher
frequency, $\nu_c=1/T_c=200$ GHz, the deviation becomes significant
and the diffusion equation is not applicable. Moreover, the damping
length of $\vec{\jmath}_s^{\bot}$ decreases with frequency due to the
`skin' effect.~\cite{Zhu08} We can conclude that the spin diffusion
equation is applicable only in the low frequency range and amounts to
an adiabatic approximation: the external perturbation is assumed to be
much slower than the internal dynamics of the electronic system.

\begin{figure}
\includegraphics[width=0.4\textwidth]{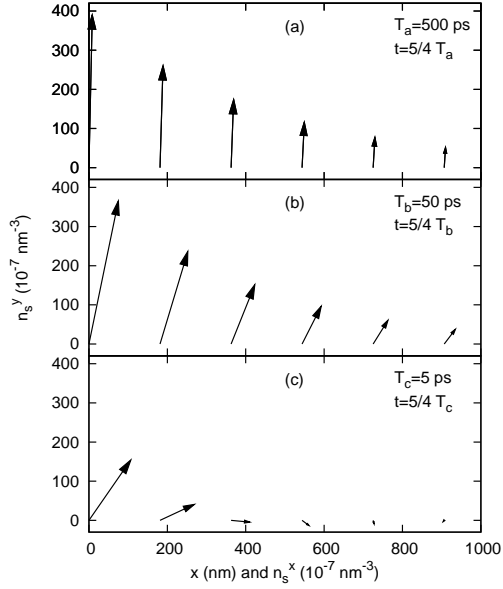}
\caption{Snapshots of the spin density $\vec{n}_s^{\bot}$ for the
same parameters as in Fig.~\ref{fig1}. \label{fig2}}
\end{figure}

Figure \ref{fig2} shows snapshots of the spin density
$\vec{n}_s^{\bot}$ for the same parameters as in Fig.~\ref{fig1}.
The spin density $\vec{n}_s^{\bot}$ deviates from $y$-axis at
$x=0$ and is noncollinear with $\vec{\jmath}_s^{\bot}$ at $x>0$ at all
of the three frequencies. This feature is different from the result
of the diffusion equation, where $\vec{\jmath}_s^{\bot}$ and
$\vec{n}_s^{\bot}$ are collinear.~\cite{Bra02,Tser02b} The phase
shift and the amplitude of $\vec{n}_s^{\bot}$ also vary with
frequency. Moreover, the damping length of $\vec{n}_s^{\bot}$
decreases with frequency again due to the `skin' effect.

According to Eq.~(\ref{critical}), the diffusion character is dominant at
the frequencies considered so far, because they are still much smaller
than the critical frequency $\nu_{\mathrm{crit}}$. This conclusion is
supported by the numerical results presented in Figs.~\ref{fig1} and
\ref{fig2}, although Figs.~\ref{fig1} (c) and \ref{fig2} (c) have
already shown weak wavelike character. The deviation from the
diffusion equation depends largely on the frequency of the spin signal
and momentum relaxation time, which varies with material, temperature,
doping and excitation condition. In the following, we show the
numerical results for a frequency $\nu_d=8.33$\,THz, where the wave
character is significant according to Eq.~(\ref{critical}).

\begin{figure}
\includegraphics[width=.4\textwidth]{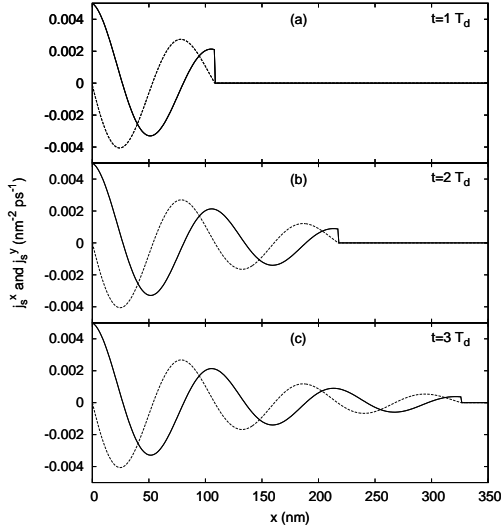}
\caption{Snapshots of $\vec{\jmath}_s^{\bot}$ at $t= T_d$, $2 T_d$,
and $3 T_d$, where $T_d=0.12$\,ps. The solid (dashed) curve is for
$j_s^x$ ($j_s^y$). \label{fig3}}
\end{figure}

\begin{figure}
\includegraphics[width=0.4\textwidth]{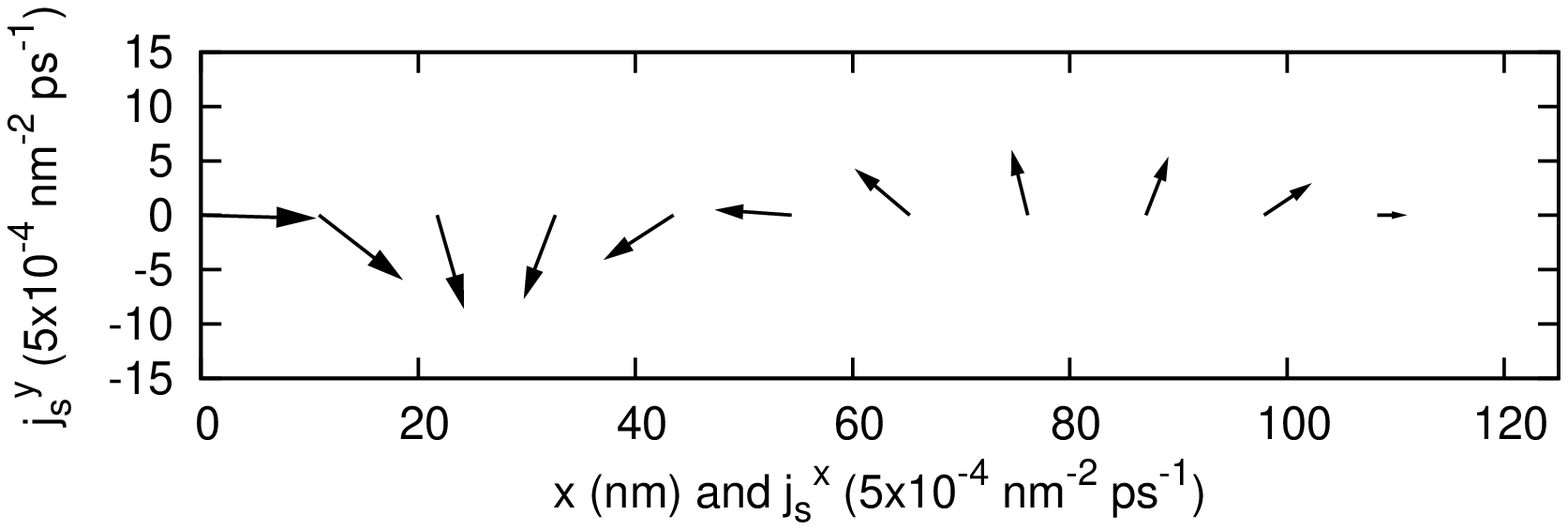}
\caption{Snapshot of $\vec{\jmath}_s^{\bot}$ (plotted as vector) at
$t=1\ T_d$. \label{fig4}}
\end{figure}

Figure \ref{fig3} shows snapshots of the spin current density
$\vec{\jmath}_s^{\bot}$ at $t=1 T_d$, $2 T_d$, and $3 T_d$,
respectively. The wave form and wave front are clearly visible in
Fig.~\ref{fig3}. The propagation velocity of the spin signal can be
estimated by tracking the motion of the wave front. The result is
approximately equal to the analytical result $c_s=906$\,nm/ps. The
phase velocity can also be estimated by measuring the wavelength
$\lambda$ and using $v_p=\lambda/T_d$. The result is roughly equal to
the wave front velocity $c_s$, which also indicates the significance
of the wave character, albeit on the length scale of the damping
length (dynamical spin diffusion length). To demonstrate the wave
character more directly, we plot the results of Fig.~\ref{fig3} (a)
again in Fig.~\ref{fig4}, where $\vec{\jmath}_s^{\bot}$ is shown in a
vector plot. Note that $\nu_d$ is beyond the frequency range in which
Eq.~(\ref{pump}) is valid, because Eq.~(\ref{pump}) is only applicable
in the adiabatic limit, $\nu\ll1/\tau$.~\cite{Tser02} Unfortunately,
there is no corresponding theoretical result for the nonadiabatic
spin-pumping in the literature. However, it is a reasonable guess that
the pumped spin current density in the nonadiabatic regime preserves
the basic feature of Eq.~(\ref{pump}): $\vec{\jmath}_s^{\mathrm{pump}}$
rotates with a certain fixed frequency. Therefore, the spin current
predicted by our results should be at least qualitatively accurate in
this frequency range.

\section{Summary\label{sec5}}

We showed that time-dependent noncollinear spin transport exhibits a
wave character for modulation of the spin current on timescales shorter
than an inverse critical frequency. A finite propagation velocity for
the spin signal can be defined due to this wave character. The spin
diffusion equation is recovered only for modulation with frequencies
less than the critical fequency, and amounts to an adiabatic
approximation of time-dependent spin transport.

\begin{acknowledgments}
We acknowledge financial support from the state of Rheinland-Pfalz
through the MATCOR program and a CPU-time grant from the John von
Neumann Institut for Computing (NIC) at the Forschungszentrum
J\"{u}lich.
\end{acknowledgments}

\appendix

\section{\label{derivation}Derivation}

Equations (6) and (7) of Ref.~\onlinecite{Qi03} are derived using
the ``mean field'' approximation
\begin{equation}
\label{mean-field}
\sum_{\vec{v}}v_x^2(\partial\hat{\rho}/\partial{x})\approx
\overline{v_x^2}\sum_{\vec{v}}(\partial\hat{\rho}/\partial{x}) .
\end{equation}
Here we show that $\overline{v_x^2}=c_s^2$ by evaluating the sums
occurring in Eq.~\eqref{mean-field}.  We start with the LHS, which we
denote by $I_1=\sum_{\vec{v}}v_x^2(\partial\hat{\rho}/\partial{x})$.
Due to the cylindrical symmetry of the system around the $x$ axis in
velocity space, $\hat{\rho}$ can be expanded in Legendre polynomials of
$u=\cos\theta$, where $\theta$ is the angle between $\vec{v}$ and
the $x$ axis, as
\begin{equation}
\hat{\rho}=\sum_{n=0}^{\infty}\hat{\rho}_n(v,x)P_{n}(u).
\label{Legendre}
\end{equation}
Transforming the
summation into an integral, we have
\begin{equation}\label{current5}
I_1=\frac{2\pi{V}m^{\ast3}}{h^{3}}\int_{-1}^{1}duu^{2}\int_{0}^{\infty}dvv^{4}
\sum_{n=0}^{\infty}\frac{\partial}{\partial{x}}\hat{\rho}_n(v,x)P_{n}(u).
\end{equation}
Using $u^2=[2P_2(u)+P_0(u)]/3$, we write the integral as
\begin{equation}
I_1=\frac{2\pi{V}m^{\ast3}}{h^{3}}\int_{0}^{\infty}dvv^{4}
\sum_{n=0}^{\infty}\frac{\partial}{\partial{x}}\hat{\rho}_n(v,x)
\int_{-1}^{1}du\frac{1}{3}\left[2P_{2}(u)
+P_{0}(u)\right]P_{n}(u).
\end{equation}
%&=\frac{2\pi{V}m^{\ast3}}{h^{3}}\int_{0}^{\infty}dvv^{4}
%\sum_{n=0}^{\infty}\frac{\partial}{\partial{x}}\hat{\rho}_n(v,x)\left[\frac{4}{15}\delta_{n,2}
%+\frac{2}{3}\delta_{n,0}\right]\\
Making use of the orthogonality relation of Legendre polynomials, we
have
\begin{equation}
I_1=\frac{2\pi{V}m^{\ast3}}{h^{3}}\int_{0}^{\infty}dvv^{4}
\frac{\partial}{\partial{x}}\left[\frac{4}{15}\hat{\rho}_2(v,x)
+\frac{2}{3}\hat{\rho}_0(v,x)\right].
\end{equation}
If the system is weakly anisotropic, we can neglect the second-order
term $\hat{\rho}_2(v,x)$,
\begin{equation}
I_1\approx\frac{4\pi{V}m^{\ast3}}{3h^{3}}\int_{0}^{\infty}dvv^{4}
\frac{\partial}{\partial{x}}\hat{\rho}_0(v,x).
\end{equation}
This approximation is consistent with Ref.~\onlinecite{vf}, where
the second-order term of the Legendre polynomials is neglected and
it is shown that this is valid if $\sqrt{\tau/(2T_1)}\ll1$.

Because $\partial\hat{\rho}_0(v,x)/\partial{x}$ is zero unless $v$
falls in a small region $[v_{F}-\Delta{v},v_{F}+\Delta{v}]$ around
the Fermi velocity $v_F$ of a system with a degenerate electron gas,
we have approximately
\begin{equation}
I_1 =\frac{4\pi{V}m^{\ast3}}{h^{3}}
\frac{v_{F}^{2}}{3}\int_{v_{F}-\Delta{v}}^{v_{F}+\Delta{v}}dvv^{2}
\frac{\partial}{\partial{x}}\hat{\rho}_0(v,x) 
 = \frac{v_{F}^{2}}{3}\, \frac{4\pi{V}m^{\ast3}}{h^{3}}\int_{0}
^{\infty}dvv^{2}\frac{\partial}{\partial{x}}\hat{\rho}_0(v,x)\ .
\end{equation}
We now need to evaluate the RHS of Eq.~\eqref{mean-field}, which we
denote by
\begin{equation}
I_2  =\overline{v_x^2}\,\frac{Vm^{\ast3}}{h^{3}}2\pi\int_{-1}^{1}du\int_{0}
^{\infty}dvv^{2}\frac{\partial}{\partial{x}}\hat{\rho}(\vec{v},x)
  = \overline{v_x^2}\, \frac{4\pi{V}m^{\ast3}}{h^{3}}\int_{0}
^{\infty}dvv^{2}\frac{\partial}{\partial{x}}\hat{\rho}_0(v,x)\ .
\end{equation}
where, in the last line, we used that the integral over $u$ projects
the contribution of $P_0$ out of $\hat{\rho}(\vec{v},x)$. Because $I_1
= I_2$, we conclude that $\overline{v_x^2}=v_{F}^{2}/3\equiv c_s^2$.

\section{\label{numerical}Numerical solution}

The basics of our numerical method have been outlined in Appendix A4
of Ref.~\onlinecite{Zhu08}. For present calculation, it has to be
augmented by a discretized version of the boundary condition on at
ferromagnet/nonmagnet interface,
\begin{equation}
\left(\Delta{t}/T_{1}+2\right)n_{s,i}^{+,l+1}=
-\left(\Delta{t}/T_{1}-2\right)n_{s,i+1}^{+,l}
+c_s^{-1}\left(\Delta{t}/\tau-2\right)j_{s,i+1}^{+,l}
+c_s^{-1}\left(\Delta{t}/\tau+2\right)j_{s,i}^{+,l+1}
\end{equation}
where the subscripts~$i$ and superscripts~$l$ stand for the discrete
space-time points, and $\Delta t$ is the numerical time step.

%\bibliography{reference_ncl}% Produces the bibliography via BibTeX.

\end{document}